\documentclass[11pt]{article}
\usepackage{graphicx}
\usepackage{amsmath}
\usepackage{amssymb}
\usepackage{amsxtra}
\usepackage{changebar}
\usepackage{placeins}

\newcommand{\bc}{\begin{center}}
\newcommand{\ec}{\end{center}}
\newcommand{\bd}{\begin{displaymath}}
\newcommand{\ed}{\end{displaymath}}
\newcommand{\be}{\begin{equation}}
\newcommand{\ee}{\end{equation}}
\newcommand{\ba}{\begin{array}}
\newcommand{\ea}{\end{array}}
\newcommand{\bea}{\begin{eqnarray}}
\newcommand{\eea}{\end{eqnarray}}
\newcommand{\bt}{\begin{tabular}}
\newcommand{\et}{\end{tabular}}

\newcommand{\bp}{\begin{picture}}
\newcommand{\ep}{\end{picture}}
\newcommand{\bfi}{\begin{figure}}
\newcommand{\efi}{\end{figure}}
\begin{document}


\title{{\huge \bf Dimension Four Wins 
the Same Game as the Standard Model Group
  }}

\author{
H.B.~Nielsen
\footnote{\large\, hbech@nbi.dk, 
hbechnbi@gmail.com}
\\[5mm]
\itshape{
The Niels Bohr Institute, Copenhagen,
Denmark}\\
}

\date{}

\maketitle

\begin{abstract}
In a previous article Don Bennett and I 
looked for, found  and proposed
a game in which the Standard Model Gauge 
{\em Group} $S(U(2) \times U(3))$ gets 
singled 
out as the ``winner''. 
This ``game'' means that the by Nature
chosen gauge group should be just the 
one, that has the maximal value for a 
quantity, which is a modification of the 
ratio of the quadratic Casimir for the 
adjoint representation and  that for
a ``smallest'' faithful representation. 
Here I propose to
extend this ``game'' to construct a 
corresponding game between different 
potential dimensions for space-time.
The idea is to formulate, how the same
competition as the one between the 
potential gauge groups would run 
out, if restricted to the potential
Lorentz or Poincare groups achievable 
for different dimensions of space-time
$d$. The remarkable point is, that 
it is the experimental space-time 
{\em dimension 4, which wins}. So the same 
function defined over Lie groups seems
to single out {\em both} the gauge group
{\em and} the dimension of space time in 
nature. 
This seems a rather strange coincidence,
unless there really is some similar 
physical reason behind causing our 
game-variable (or goal variable) to be 
selected to be 
maximal. It has crudely to do with 
that the groups preferred are easily 
represented on very ``small'' but yet 
faithful representations.     
 \end{abstract}

\section{Introduction}
The main idea of the present series 
of articles  is to seek some game, that
at the same time can select out the gauge 
group observed in nature - let us suppose
we should have the Standard Model 
Group $S(U(2)\times U(3))$ in nature, say -
and {\em also} the gauge group (whatever 
that
means) of the (gravitational) general 
relativity by saying that nature has 
chosen the ``winner'' in this game. 
That is to say we look for a 
group-characteristic quantity (``goal 
quantiy'') which happens to  be largest 
possible for  both the gauge group of the 
Standard Model and a group associate 
with the Lorentz transformations
(or somehow with the gauge transformations 
in general relativity), the 
Lorentz group say. We could then claim 
that such a goal quanity specifies both 
the gauge group for the Standard Model say
and the Lorentz group meaning thereby the 
dimension of space time. If the quantity 
is reasonably simple, this could be an 
explanation for both the gauge group 
and the dimension of space time.
We could then  answer: Why do we have 
in nature just the Standard Model Group 
$S(U(2)\times U(3))$ and why just 4 
space time dimensions?  
In the 
previous article\cite{seeking} we sought 
in this way to invent a game or rather 
{\em a ``goal quantity''} - which were at 
first
the ratio $C_A/C_F$ of the quadratic 
Casimirs for the group in question 
for the adjoint representation to the 
quadratic Casimir of  
some ``small'' but still faithful
representation- in such a way 
  that this ratio would take its 
largest value for the by nature chosen 
(gauge) group.

Both ourselves and others have earlier 
also made other attemps to find arguments 
pointing as well the gauge group
\cite{Brene} and the dimension 
\cite{Tegmark}\cite{RD,Foerster, 
RDrev, Astri, RDDon1} \cite{Ehrenfest}
\cite{Kane}.
We shall shortly review earlier works in 
appendix \ref{earlier}.

In section \ref{review} we shall review
the previous work \cite{seeking}. 


Actually N. Brene and I had already 
earlier proposed another game that 
essentially pointed also to the Standard 
Model Gauge \underline{group} being the 
winner
\cite{Brene}, but it is the more recent 
proposal with the quadratic Casimirs   
or rather their ratio $C_A/C_F$ which we 
seek to generalize to determine the 
dimension of space time in this article. 
This concept of the 
gauge group for general relativity
may be a bit imprecise, and so I want 
at first to simplify a little bit by 
making a few ad hoc decisions to 
extract a group, that essentially is 
the gauge group of general relativity,
even if we should not have chosen the 
definition of 
this concept  completely clearly  yet.

A first candidate, which is for me 
rather attracktive for the purpose, is 
simply the Lorentz group meaning the 
group of Lorentz boosts and 
rotations.

You could consider the attitude of the 
present article and the foregoing one in 
the 
series \cite{seeking} as attempts to 
extract the information as discussed in 
the article \cite{Rugh} 
contained in the group structure of the 
Stabdard Model gauge group and the 
dimension of space-time.
 Of course the hope could be that one 
would in this way learn about the true 
theory, that might be behind the Standard 
Model by finding some regularity(as we 
may say we do in the present series of 
papers).





The reader should consider these different
proposals for a quantity to maximize
(= use as goal quantiy) as rather closely 
related versions of a quantity suggested 
by a perhaps a bit vague idea being 
improved successively by treating the 
from our point of view a bit more 
difficult to treat Abelian part (the 
translation part of the Poincare group)
at least in an approximate way. 
One should have in mind, that this somewhat
vague basic idea behind is: The group 
selected by nature is the one that 
counted
in a ``normalization determined from the 
Lie algebra of the group'' can be said to 
have a faithfull representation ($F$)
the matrices of which move as little as 
possible, when the group element being 
represented move around in the group.

Let me at least clarify a bit, what is 
meant by this statement:

We think by representations as usual
on linear representations, and thus it 
really means representation of the group 
by means of a homomorphism of the group
into a group of matrices. The requirement 
of the representation being faithful then 
means, that this group of matrices shall
actually be an isomorphic image of the 
original group. Now on a system of 
matrices we have a natural metric, 
namely the metric in which the distance 
between two matrices ${\bf A}$ and 
${\bf B}$ is given by the square root of 
the trace of the numerical square of the 
difference
\begin{equation}
dist = \sqrt{tr(({\bf A -B})
({\bf A -B})^+)}. \label{distrep}
\end{equation}  
To make a comparison of one group and 
some representation of it  with 
another group and its representation 
w.r.t. to, how fast the representation 
matrices 
move for a given motion of the group
elements, we need a normalization giving 
us a well-defined metric on the groups, 
w.r.t. which we can ask for the rate 
of variation of the representations.
In my short statement I suggested 
that this ``normalization should be 
determined from the Lie algebra of the 
group''. This is to be taken to mean 
more precisely, that one shall consider 
the {\em adjoint} representation, which
is in fact completely given by the 
Lie algebra, and then use the same 
distance concept as we just proposed for 
the matrix representation 
$ \sqrt{tr(({\bf A -B})
({\bf A -B})^+)}$. In this way the 
quantity to minimize would be the ratio
of the motion-distance in the 
representation - $F$ say - and in the 
Lie algebra representation - i.e. the 
adjoint representation. But that ratio 
is just for infinitesimal motions 
$\sqrt{C_F/C_A}$. So if we instead 
of talking about what to minimize, 
inverted it and claimed we should 
maximize we would get $\sqrt{C_A/C_F}$ 
to be {\em 
maximized}. Of course the square root 
does not matter, and we thus obtain in 
this 
way a means to look at the ratio $C_A/C_F$
as a measure for the motion of an element 
in the group
compared to the same element motion on 
the representation.

It might not really be so wild to think
that a group which can be represented in 
a way so that the representation varies 
little when the group element moves around
would be easier to get realized in nature 
than one that varies more. If one imagine 
that the potential groups become  good symmetries by 
accident, then at least it would be less 
of an accident required the less the degrees of 
freedom moves around under the to the 
group corresponding 
  symmetry (approximately). It is really 
such a 
philosophy of it being easier to get some 
groups approximately being  good 
symmetries
than other, and those with biggest 
$C_A/C_F$ should be the easiest to  
become
good symmetries by accident, I argue for. That is indeed the 
speculation behind the present article
as well as the previous one \cite{seeking} that symmetries may appear 
by accident(then perhaps being stregthened to be exact by some means
\cite{Foerster, Damgaard}).

But let us stress that you can also look 
at the present work and the previous one 
in the following phenomenological 
philosophy:

We wonder, why Nature has chosen just 4
(=3+1) 
dimensions and why Nature - at the present 
experimentally accessible scale at least - 
has chosen just the Standard Model 
group $S(U(2)\times U(3))$? Then we 
speculate that there might be some 
quantity characterizing groups, which 
measures how well they ``are suited '' 
to be the groups for Nature. And then we 
begin to {\em seek} that quantity as 
being some 
function defined on the class of abstract 
groups - i.e. 
giving a number for each abstract (Lie?) 
group -
of course by proposing for ourselves at 
least  various versions or ideas 
for what such a {\em relatively simple}
function defined on the abstract Lie 
groups could be. Then the present works 
- this paper and the previous one\cite{seeking} - 
represents the present status of the 
search: We found that with small 
variations the types of such functions
representing the spirit of the {\em little 
motion of the ``best'' faithful 
representation},i.e. essentially the 
largest $C_A/C_F$, 
turned out truly to bring Natures choices 
to be (essentially) the winners.

In this sense we may then claim that we 
have found by phenomenology that at 
least the ``direction'' of a quantity like 
$C_A/C_F$ or light modifications of it 
is a very good quantity to make up a 
``theory'' for, why we have got the 
groups we got!

\subsection{Outline}

In the following section \ref{review}
we review the main results of the 
$C_A/C_F$ quantity, which we in the 
previous article studied for the various 
Lie groups in order to discuss, that the 
Standard Model group could be made to be 
favored. In section \ref{Lorentz} we then 
extract and concentrate on those groups 
which can be Lorentz groups. The main 
content of both these sections are 
actually the tables listing the results 
of the quantities proposed to be 
maximized 
for the relevant groups. In section 
\ref{conclusion} we resume and conclude,
 that actually we may be on the track to 
have found a {\em common} reason or 
explanation for, that we have 3 +1 
dimensions, and  for the gauge group of 
the Standard Model!            

In the appendix we have put a review 
of previous attempt to argue for why 
we have just 4 dimensions.

\section{Our Previous Numbers}
\label{review}
In the previous work by D. Bennett and 
myself \cite{seeking} we essentially 
collected the 
ratios (related to Dynkin index
\cite{Dynkinindex}) $C_A/C_F$,
where we for the representation of the 
group in question $G$ selected that 
representation $F$, which would give 
the largest value for this ratio $C_A/C_F$.
(In the table we give in a few cases 
two proposals for $F$, but really it is 
what one would loosely call the smallest
faithful representation). But we shall 
have in mind that this ratio is only 
well-defined for the simple Lie groups; and
it is thus only for the simple groups we 
could make a clean table as the one just 
below. For semi-simple Lie groups 
it is strictly speaking needed to specify 
a replacement quantity, that can be the 
needed generalization to semisimple Lie 
groups. One shall naturally construct 
a logarithmic average
(see  formula (\ref{f2}) or 
(\ref{semisimple}) for what 
we mean by 
a ``logarithmic average'')
 weighted with the 
dimensions of the various simple group
factors contained in the semisimple Lie 
group written as a product of its simple 
invariant subgroups. Extending the 
generalization even to inclusion of 
$U(1)$-factors in the Lie group gets 
even a bit more arbitrary, but we did 
choose the rule of counting the $U(1)$
factors as, if they had $C_A/C_F$ equal
to unity (yes, that were at first but 
then in that article we argued for a 
correction factor derived from the 
way the elements in the center of the 
group are identified ( by division out
a subgroup), and the combined result 
of these rules became equivalent to the 
introduction of formally taking 
$C_A/C_F \rightarrow e_A^2/e_F^2$ as 
described below in item IV in subsection
\ref{subsec2c1})   . The problem with 
the $U(1)$'s,
the Abelian groups, is that the adjoint 
quadratic Casimir $C_A$ is just zero and 
does not provide a good normalization.
But although we thus have to declare 
formally $C_A/C_F$ to be unity at first 
for $U(1)$ we take the opportunity to 
- and we think it is very natural -
to include a correction depending 
not only on the Lie-{\em algebra} but 
also on the {\em group} structure 
in a way roughly describing that 
a $U(1)$ representation with a small 
``charge'' is ``smaller'' than 
one with a larger  charge
in a very similar way to the way in which 
a small quadratic Casimir signals 
a ``small'' representation. 

Here we give our (essentially Dynkin index)
ratios for the {\em simple} Lie groups:
    
{\bf Our Ratio of Adjoint to ``Simplest''
(or smallest)
Quadratic Casimirs $C_A/C_F$}
\begin{eqnarray}
\frac{C_A}{C_F}|_{A_n} & =&
\frac{2(n+1)^2}{n(n+2)} =
\frac{2(n+1)^2}{(n+1)^2 -1} =
\frac{2}{1-\frac{1}{(n+1)^2}}\\
\frac{C_A}{C_{F \; vector}}|_{B_n}& =&
\frac{2n-1}{n}= 2 - \frac{1}{n}\\
\frac{C_A}{C_{F \; spinor}}|_{B_n}&=
&\frac{2n-1}{\frac{2n^2 +n}{8}} =
\frac{16n -8}{n(2n+1)}\\
\frac{C_A}{C_F}|_{C_n} &=&
\frac{n+1}{n/2 +1/4} =
\frac{4(n+1)}{2n+1}\\
\frac{C_A}{C_{F \; vector}}|_{D_n}&=&
\frac{2(n-1)}{n-1/2}=
\frac{4(n-1)}{2n-1}\\
\frac{C_A}{C_{F \; spinor}}|_{D_n}&=&
\frac{2(n-1)}{\frac{2n^2-n}{8}}
= \frac{16(n-1)}{n(2n-1)}\\
\frac{C_A}{C_F}|_{G_2} &=& \frac{4}{2} =2\\
\frac{C_A}{C_F}|_{F_4} &=& \frac{9}{6} =
 \frac{3}{2}\\
\frac{C_A}{C_F}|_{E_6} &=&
\frac{12}{\frac{26}{3}} = \frac{18}{13}\\
\frac{C_A}{C_F}|_{E_7}&=&
\frac{18}{\frac{57}{4}} = \frac{72}{57}
= \frac{24}{19}\\
\frac{C_A}{C_F}|_{E_8}&=& \frac{30}{30} =1
\label{ratiotable}
\end{eqnarray}
For calculation of this table seek help
in\cite{Rittenberg, MacFarlaine}

In the just above table we have of course used the conventional 
notation for the classification of Lie algebras, wherein the 
index $n$ on the capital letter denotes the rank of the Lie algebra, and:
\begin{itemize}
\item $A_N$ is $SU(n+1)$,
\item $B_n$ is the odd dimension 
orthogonal group  Lie 
algebra for $SO(2n+1)$ or for its 
covering group Spin(2n+1),
\item $C_n$ are the symplectic Lie 
algebras.
\item $D_n$ is the even dimension 
orthogonal Lie algebra for $SO(2n)$ or 
its covering group $Spin(2n)$,
\item while $F_4$, $G_2$, and $E_n$ for $n=6,7,8$ are the exceptional Lie 
algebras.  
\end{itemize}

The words $spinor$ or $vector$ following in the index the letter $F$
which itself denotes the ``small'' representation - i.e. most promissing 
for giving a small quadratic Casimir $C_F$ - means that we have 
used $F$ respectively the sinor and vector representation. 

It may be reasuring to check that our goal quantity for the simple 
groups $C_A/C_F$ becomes the same for the cases of isomorphic Lie algebras
such as $B_2 \cong C_2$ and 
$C_3 \cong A_3$.

\subsection{Development of the 
Gauge Group
Determination Proposal}
\label{subsec2c1}
It may be best to describe the proposal
for the quantity to be maximized for the 
gauge group by describing how a 
phenomenological discussion adjusting 
small problems can be guided towards
the  final rule. Let me review the 
work \cite{seeking} as a successive 
discussion of larger and larger 
classes of groups towards finding a goal 
quantity 
that would make the Standard Model group
win the game of making it maximal. 

But just before that I want to be a bit 
concrete and present the typical type of group
that we at all consider a possibility as a
gauge group.Indeed we  imagine that it can be 
written as a cross product of 
Lie groups with at the end some subgroup
of the center being divided out.:

We were all over in the paper 
\cite{seeking}
satisfied  in 
practice by considering  
a Lie group of the form of a cross product
of some $U(1)$ groups and some simple 
Lie groups finally modified by dividing
out some discrete  subgroup of the 
center. That is to say we have in mind 
groups of the form
\begin{equation}
G =\left ( U(1)\times U(1) \times \cdots 
\times 
U(1) \times SU(2)(say)\times \cdots ...\times 
G_{max}\right ) /D
\end{equation}   
where the $\times$-product runs over a 
number of occurrences of all the possible 
simple Lie groups as classified say by
their Dynkin diagrams up to the last one
for the group $G$ in question here denoted
$G_{max}$. We imagine using the covering 
groups for the Lie algebras in question 
and take only the compact groups. 
Finally then the {\em group}
rather than Lie algebra structure is 
achieved by dividing some discrete 
subgroup $D$ out of the center of 
the group achieved by the cross product
without modification. 
This division out of a subgroup from 
the center only has significance for 
the {\em group} but not for the Lie 
algebra. The gauge fields in a gauge 
field theory a priori only depends on the
Lie {\em algebra}, but we as a very 
important point in our works assign a physical
significance to even the Lie {\em group}
by making use of that the {\em group} 
structure restricts the representations
that are allowed. Thus one can in a 
phenomenological way read off the 
{\em group} structure (and thereby what
were divided out) by studying the 
representations occurring as 
representations for the matter fields.
It is e.g. the empirical charge 
quantization rule 
\begin{equation}
y/2 + I_W + ``triality''/3 = 0 
\hbox{(mod 1)}
\end{equation}
where $I_W$ is the weak isospin and 
$y$ the hypercharge, and it tells that 
the electric charge $Q = T_{3W}+y/2$ 
is integer for particles with zero 
triality, written as ``triality'', and 
becomes 1/3 modulo unity for triality 
1(mod3), while -1/3(mod 1) for ``triality''
=2 (mod 3).   

The reader should 
have in mind that the typical covering 
groups as e.g. $SU(N)=A_{N-1}$ has often
a nontrivial center with a finite number
of elements. E.g. $SU(N)$ has a center 
isomorphic to the group $Z_N$ of the 
integer numbers counted modulo $N$. 
The whole center in the cross product 
thus becomes the cross product of the 
typically discrete centers for the simple 
Lie groups crossed further with the 
$U(1)$'s which each of them are all 
center (since they are Abelian).
For example the center of the cross 
product 
$U(1) \times SU(2) \times SU(3)$ that 
shall be used to produce the Standard 
Model group (and which has the Standard 
Model Lie algebra) is $U(1)\times 
Z_2\times Z_3$. By dividing out 
of this a cleverly chosen with $Z_6$
isomorphic (discrete) subgroup 
$D$ generated by the element $(2\pi, 
1 (mod 2), 1 (mod 3))$ (where 
``$1 (mod 3)$'' e.g. the element in 
$Z_3$ being the class of integers with 
rest 1 modulo 3)   we obtain 
a {\em group} which does not have 
as representations of the group all the 
representations of the cross product
$U(1) \times SU(2) \times SU(3)$
or better  its covering group, but
nevertheless those representations 
occurring for particles in the Standard 
Model.

The following successive proposals 
are then made for larger and 
larger subsets of the groups 
of the type considered, beginning
in I with the simple Lie groups,
then in II the semisimple etc:

\begin{itemize}
\item{\bf{I}} The ground idea for the goal
quantity is the ratio $C_A/C_F$ in which 
the symbols $C_A$ and $C_F$ are the 
quadratic Casimirs for the group 
in question for respectively the 
adjoint representation $A$ and another 
representation $F$, which then in the 
search for a maximal ratio $C_A/C_F$
will lead to choosing $F$ with minimal 
$C_F$. To avoid that $F$ be the  trivial 
representation we shall what is very 
reasonable we would say require $F$ to
be faithful. 

This simple starting proposal $C_A/C_F$
for defining a ``goal quantity'' to seek
maximum for is really only working for
{\em simple non-Abelian Lie groups}.
(so for other Lie groups it shall need
some improvements to be a good and 
weldefined quantity)

In fact  the reader shall have in mind 
that 
\begin{itemize}
\item{I.1}
The ratio $C_A/C_F$ does not suffer 
from the normalization problem of the 
generators representing the Lie algebra,
because we take the ratio so that scaling
the convention for the Lie algebra basis, 
if changed will change the numerator 
$C_A$ and the denominator $C_F$ by the 
same
factor.

\item{I.2} But at first even this {\em 
ratio} $C_A/C_F$ is only weldefined for a 
simple Lie group. In the case of even 
a still semi-simple Lie group there is 
namely an ambiguity  in the 
normalization of the basis vectors for 
the Lie algebra of one of the simple 
components relative to another simple 
component. So just dividing two Casimirs
is not sufficient to make a normalizaton
convention independent ratio, as it were 
in the simple Lie algebra case. 

\item{I.3} If we do not specify the 
normalizations of the basis vectors and 
thereby their representations, then we
get a notation dependent quantity for the 
(quadratic) Casimir operators and thus 
the quadratic Casimirs.

\item{I.4} If we have $U(1)$ as factors 
in the $\times$-product, then we have the 
obvious trouble with our first proposal
$C_A/C_F$ that the adjoint representation
of $U(1)$ is trivial, or rather the 
Abelian Lie group $U(1)$ has no 
(meaningful) adjoint representation and
thus $C_A$ becomes meaningless for 
$U(1)$.     
\end{itemize}

\item{\bf{II}} Next let us improve the first 
proposal $C_A/C_F$ for the goal quantity,
by generalizing it in a good way to 
the {\em semi-simple Lie groups}.

The problem which we first have to solve 
in 
extending in a meaningfull way the 
proposed quantity is to even ignoring 
at first the $U(1)$ groups and restricting
ourselves to semi-simple groups find some 
way of defining a quantity like the 
for the simple Lie group or algebra 
well-defined quantity $C_A/C_F$.

Since for a simple group the ratio 
$C_A/C_F$ {\em is} welldefined the 
obvious idea to make analogous expression
for a semi-simple, which is just a 
$\times$-product of several simple 
Lie groups $S_1 \times S_2 \times \cdots
S_n$, is taking some sort of average 
over the separate simple groups $S_i$
of the quantities $C_A/C_F|_{S_i}$ for 
the various simple groups. The proposal 
which we thought were reasonable was to 
average logarithmically (see (\ref{f2})
or (\ref{semisimple}) to get an idea what 
``logarithmic averaging''
means) and weighting with
the dimension of the Lie groups. That 
means we proposed to use as the average 
that should replace the $C_A/C_F$ for 
the simple group in the semi-simple case
\begin{eqnarray}
``C_A/C_F \hbox{replacement for} S_1 
\times S_2 \times \cdots \times S_n''\hbox
{            }&
&\\
= (C_A/C_F|_{S_1})^{\frac{dim(S_1)}{\sum_i 
dim(S_i)}} *(C_A/C_F|_{S_2})^{\frac{dim(S_2)}
{\sum_i dim(S_i)}}* \cdots *
(C_A/C_F|_{S_n})^{\frac{dim(S_n)}{\sum_i 
dim(S_i)}}&&.\label{semisimple} 
\end{eqnarray}       
It is of course a priori an ad hoc 
choice to weight just with the dimensions
 $dim(S_i)$  of the simple groups $S_i$,
but is an extremely natural choice, but 
of course the type of choice we could 
revise, if we should look for some 
little adjustment of our proposal to make 
it be more successful.

You may consider the quadratic Casimir
as representing a metric tensor describing
distances in the representation space 
for coordinates originating from 
the Lie algebra or Lie group. If one would
think of the volume (of dimension $dim(G)$
of course)
for a faithful representation, $F$ say, 
relative to the adjoint representation 
volume, it would be $(C_F/C_A)^{dim(G)/2}$.
So you could see our quantity 
$``C_A/C_F \hbox{replacement for} S_1 
\times S_2 \times \cdots \times S_n''$
as being the $dim(S_1\times S_2 \times
\cdots \times S_n)/2$'th root of the 
volume ratio of the adjoint representation
$A$, i.e. $Vol(A)$ relative to that 
of the faithful 
representation $F$, i.e. $Vol(F)$,
\begin{equation}
``C_A/C_F \hbox{replacement for} S_1 
\times S_2 \times \cdots \times S_n''
= (\frac{Vol(A)}{Vol(F)})^
{\frac{1}{dim(G)}}.\label{root}
\end{equation}
This simple and nice interpretation
supports estetically the use of the 
dimensionality of the various simple 
Lie groups being used to weight  the 
logarithmic average. We can namely 
instead of talking at first about the 
quadratic Casimirs say that we talk about 
the volume ratio of the adjoint 
representation and the representation 
$F$ from the start. The ratio 
of the volumes of two representations
in the natural metric defined above 
in (\ref{distrep}) is a very simple and 
beautiful quantity. We then take the 
root of (half) the dimension
of the group to make it depend on the 
structure of the various simple subgroups
rather than on the total number of them 
or their dimension in a too strong way.
By using this root choice (\ref{root})
we obtain the good feature, that you 
cannot obtain the large quantity just by
taking a group with a high dimension 
by taking for instance a cross product
of a lot of groups. For single simple 
groups we also achieve that our root
quantity becomes just the $C_A/C_F$,
from which we started. And you can even 
see from the table above that in the 
limit of the rank going to infinity 
the various series of infinitely many
simple Lie groups have our quantity go
nicely to 2.\footnote{According to our 
rule one shall choose the representation 
$F$ to be the faithful representation 
making $C_A/C_F$ maximal for the given 
group. Whether to choose the spinor 
or vector possibility for $F$ according 
to this rule will shift for the 
$Spin(N)\approx SO(N)$
groups meaning the $B_n$ and $D_n$  
at $Spin(8)\approx SO(8)$ for which the 
spinor and the 
vector representations are isomorphic. 
For the assymptotic case of large ranks
$n$ shall one as $F$ use the vector
possibility, and that gives the limit 
2. } In this way we get a very 
{\em balanced} quantity, not favoring
at first neither large nor small 
dimensions for the group dramatically.

Hereby we think, we have proposed a 
very nice and beautiful quantity for 
the semi-simple groups.

\item{\bf {III}} Next we have the problem 
with groups having $U(1)$-factors:

For the $U(1)$ our $C_A/C_F$ hardly 
makes any sense, and so we have to invent
a replacement essentially arbitrarily.
In order to do that let us have in mind 
that whenever our quantity $C_A/C_F$ 
makes sense it is for trivial reasons 
always bigger than unity. We could 
namely always as a special possibilty 
for the faithful representation $F$ 
use the adjoint representation itself,
in which case of course the ratio
$C_A/C_F = C_A/C_A =1$. So since we 
shall choose the representation $F$ 
so as to maximize the ratio $C_A/C_F$,
it must always be larger than or equal
to this possibility value $1$. 

When we so to speak shall invent a 
value for the replacement of the 
$C_A/C_F$ for the Abelian group $U(1)$,
we must 
in order not to violate the 
trivial lower bound at least choose the 
value larger than or equal to $1$.

Since all representations of an Abelian
group are one-dimensional, there is w.r.t.
dimension only one representation and 
thus only one choice for $F$. We therefore
ar first proposed to replace $C_A/C_F$ by
the lowest for trivial reasons value $1$.
However, truly an Abelian group $U(1)$
has a series of different representations
given by a ``charge'' $e$.  

In other words we propose to choose:

\begin{equation}
``C_A/C_F\hbox{replacement for} U(1)''
=1. 
\end{equation}
   
It is of course then the obvious 
generalization to the groups being cross
products of $U(1)$'s with a semi-simple 
group that we shall average  this $1$ 
logarithmically with the $C_A/C_F$'s 
for the simple groups weighting as above
with the dimension of the Lie groups.

This means that we have now come to the
proposal:
\begin{eqnarray}
``C_A/C_F\hbox{replacement for} U(1) \times
\dots \times U(1)\times S_1 \times \cdots
\times S_n ''&&\\
= 1^{\frac{1}{\# U(1)'s + \sum_i dim(S_i)}}
* \cdots * 
1^{\frac{1}{\# U(1)'s + \sum_i dim(S_i)}}&&\\
* (C_A/C_F|_{S_1})^{\frac{dim(S_1)}
{\#U(1)'s+\sum_i 
dim(S_i)}}* \cdots *(C_A/C_F|_{S_n})^
{\frac{dim(S_n)}{\#U(1)'s + \sum_i S_i}}
&&\\
= (C_A/C_F|_{S_1})^{\frac{dim(S_1)}
{\#U(1)'s+\sum_i 
dim(S_i)}}* \cdots *(C_A/C_F|_{S_n})^
{\frac{dim(S_n)}{\#U(1)'s + \sum_i S_i}}.
&&   
\end{eqnarray} 
     Here $\#U(1)'s$ means the number
of $U(1)$-factors in the $\times$-product
forming the group $G$ under 
study/evaluation w.r.t. the goal quantity.
Of course the full dimension of the 
Lie group $G$ is just 
\begin{equation}
dim(G) = \#U(1)'s + \sum_i dim(S_i).
\end{equation}

At this stage it will of course {\em not}
w.r.t. maximizing the goal quantiy 
$``C_A/C_F\hbox{replacement for} U(1) \times
\dots \times U(1)\times S_1 \times \cdots
\times S_n ''$ pay to have any $U(1)$'s 
at all. So we have at this stage in 
developping the goal quantiy no chanse to
make a group that like the Standard Model
has an invariant Abelian subgroup 
$U(1)$ have any chance of winning the 
game of maximization.

\item{\bf{IV}} Improvement for Abelian and 
the {\em group} structure:

In order to make it at all pay for the 
maximization to have at least some factor
$U(1)$ like the Standard Model happens 
to have, we must open for the possibility
of letting a $U(1)$ invariant subgroup
contribute more than just the absolute 
minimum
$1$ as replacement for its meaningless
$C_A/C_F$. But the irreducible 
representations of the 
$U(1)$ are all just one-dimensional
representations with the single element
in the unitary matrix being just a phase 
factor $exp(i e \delta)$, where $\delta$
is the phase describing the element in 
$U(1)$ and $e$ is a ``charge'' for the 
representation in question. It is wellknown
that the various representations of $U(1)$
are characterized by such ``charges'' $e$.
The quadratic Casimirs are given as the 
square of the ``charges'' $ C_R 
\rightarrow e_R^2$, where $e_R$ is the 
``charge'' for the representation $R$.

Now let us have in mind that for our 
purpose studying gauge groups we mainly
have in mind the charges for various 
particles, and that when we give a 
physical meaning to the gauge {\em group}
rather than just to the gauge Lie algebra 
we do that on the basis of 
phenomenology suggested  restrictions on
the representations/particles occurring 
in the model. We can then ask the 
questions: What is the lowest non-zero
charge $e_A$ say on a particle in 
accordance with the restrictions from the 
{\em group}-structure, when this particle
has no non-abelian transformations(i.e.
when it transforms trivially/i.e. not at 
all under the other factors in the group ?
And we can also ask what is the absolutely 
(numerically) smallest ``charge'' $e_F$
say on any particle allowed under the 
group rule(whatever its couplings to the
non-Abelian Lie groups might be)?
The indexes suggested here were chosen 
to be suggested  to used   to form a ``replacement '' for 
the $C_A/C_F$ for a $U(1)$ being instead 
of the at first proposed $1$ now improved
to $e_A^2/e_F^2$. We can only obtain this
ratio $e_A^2/e_F^2$ to be different from 
unity by having a gauge group obtained 
by dividing out a discrete subgroup of the
center of the starting pure cross product.
Actually this choice is very reasonable 
in as far as the $e_F$ charge is the 
smallest possible non-trivial charge 
quite analogous to our in the non-abelian 
case choice of the representation $F$
being the smallest faithful one. So the 
only a little bit ad hoc choice is to
replace the adjoint representation for 
the non-abelian case by the smallest 
representation of the $U(1)$ that does 
{\em not} mix up with non-abelian groups.
We think this is pretty much the simplest
reasonable replacement. 

Choosing this procedure we get to the 
final proposal:
\begin{eqnarray}
``C_A/C_F\hbox{replacement for} U(1)^{(1)} \times
\dots \times U(1)^{(m)}\times S_1 \times 
\cdots
\times S_n ''= &&\\
= \left ( \frac{e_A^{(1)2}}{e_F^{(1)2}} *
\cdots \frac{e_A^{(m)2}}{e_F^{(m)2}}*
\frac{C_A}{C_F}|_{S_1}^{dim(S_1)}* \cdots
*\frac{C_A}{C_F}|_{S_n}^{dim(S_n)} \right )^
{\frac{1}{\#U(1)'s + \sum_i dim(S_i)}}.
\label{lastformula}       \end{eqnarray}    
Here of course $m=\#U(1)'s$ and the index 
in the round brackets on the charges 
enumerates the various $U(1)$ factors in 
the cross product. It should be had in 
mind that the nontrivial (i.e. not just 
$1$) ratios $\frac{e_A^{(i)2}}{e_F^{(i)2}}$
only come in play when a discrete subgroup
of the center of the cross product have 
been divided out. Remember that I 
in reviewing the work \cite{Brene} 
told that this division out of a discrete
subgroup of the center were in the 
``skewness'' estimation \cite{Brene} an 
important ingredient in reducing the 
symmetry and thus got favoured by asking 
for ``skewness''. So letting this 
division out be favored in the game 
gives a favoring of the Standard Model
{\em group} which has relatively much 
such ``division out''. This is how 
we got in this last step  
introduced a favoring of the  
``division out'' although we started out
with a $C_A/C_F$ ratio which were purely
dependent on the Lie algebra.  
\end{itemize}

\subsection{Standard Model Group 
Wins}
Using the table (\ref{ratiotable}) 
inserting it into (\ref{lastformula})
we may now contemplate which group should
win the ``game'' of obtaining the largest 
value for the goal quantity 
(\ref{lastformula}). 

First we see from the table 
(\ref{ratiotable}) that the groups 
favored even in the system that were made 
balanced in such a way that the ratio 
$C_A/C_F$ goes to a constant -actually 2 -
for very large ranks $r \rightarrow 
\infty$ are the small rank ones:
$SU(2) = A_1$ is the winner among the 
simple group with its  $C_A/C_F
=\frac{2}{1-\frac{1}{(n+1)^2}}$ for
$A_n$ which for $n=1$ gives $8/3$.

The next among the simplest Lie groups
is $SO(5) = B_2 = C_2$ with 
$C_A/C_F= 12/5$, which is obtained by 
using the spinor representation for 
$B_2$ giving $\frac{16n -8}{n(2n+1)}|_{n=2}
= \frac{12}{5}$, while the vector representation gives the less competitive 
$C_A/C_F|_{B_2 \; vector} = 3/5$, and 
we can check that the isomorphic $C_2$ also
gives as it should $12/5$. 
First at 
the third place we find the Lie group
$SU(3) =A_2$ with its $C_A/C_F|_{A_2} =
9/4$ the group we would hope to win 
over the $SO(5)$, because it is 
$SU(3)$ and not $SO(5)$ which occurs 
in the Standard Model. 

But now in the competition between 
the $SO(5)$ and the $SU(3)$ comes in 
some help for $SU(3)$ in our final 
proposal:

\begin{itemize}
\item{{\bf Dimension of $SU(3)$ is lower 
than that of $SO(5)$.}}
Comparing the semisimple groups formed 
by crossing $SU(2)$ with respectively
$SO(5) $ and $SU(3)$ we obtain the 
$C_A/C_F$ ratios when weighted according 
to (\ref{semisimple}) to be
\begin{eqnarray}
``\frac{C_A}{C_F} replacement''|_{SU(3)
\times SO(5)} & = & (\frac{8}{3})^{3/13}
*(\frac{12}{5})^{10/13} =2.459068704\\
 ``\frac{C_A}{C_F} replacement''|_{SU(3)
\times SU(3)} & = & (\frac{8}{3})^{3/11}
*(\frac{9}{4})^{8/11} = 2.356709384
\label{exa}  \\
\end{eqnarray}
The $(\frac{12}{5} /\frac{9}{4} -1) *100
\% = 6.6666667 \% $ higher value for 
$SO(5)$ over $SU(3)$ is by the 
logarithmic dimensional weighting reduced 
to $(\frac{  2.459068704}{ 2.356709384}
-1)*100\% = 4.343315332\%$.

\item{{\bf Involving a $U(1)$ and the 
division out of a central subgroup.}}

According to the details of the definition
of our ``goal quantity'' when involving 
$U(1)$ cross product factors we have the 
possibility of obtaining the 
dimensional'th 
root of factors $e_A^2/e_F^2$ from 
(\ref{lastformula}). As by our definitonal
choice above the difference between the 
charge $e_A$ and $e_F$ is that $e_A$ 
should be for representation {\em with
only the $U(1)$ charge but trivial
under the  non-Abelian groups} while 
$e_F$ can be chosen for any faithfull
representation, the ratio $e_A/e_F$ can 
only be bigger than unity by involving 
a rule for allowed representations 
of the {\em group} (rather than just 
the Lie algebra). That is to say we
need to involve the center so to speak 
of one of the covering groups of the 
non-Abelian Lie groups. We can then 
e.g. for $SU(3)$ which has a center 
isomorphic to the group of integers modulo
3, i.e. $Z_3$, obtain a divison out of 
a $Z_3$ and get a ratio of 3 for 
$e_A$ over $e_F$ if we wish. The reader 
should check that using our formula 
(\ref{lastformula}) for the goal quantity 
and imagining various {\em groups} 
obtained by various divison-outs 
of the center we can with only {\em one}
$U(1)$ cross product factor only divide 
out say $Z_2$ once if we want to get 
the effect of this division out getting 
felt for the ratios $e_A/e_F$ obtainable,
whereas we can even with only one $U(1)$
mannage to get both $Z_2$ and $Z_3$ give
rise to effective factors of the type 
$(e_A/e_F)^2$. With just one $U(1)$ factor
we can thus gain a factor 
$(e_A^2/e_F^2)^{1/dim(G)}$ 
which gives $(1/3^2)^{1/dim(G)}$ for 
$SU(3)$ and $(1/2^2)^{1/dim(G)}$ for $SU(2)$
in the goal 
quantity (where $dim(G)$ is the 
dimension of the full group). However, 
once we have already gotten such a gain 
from $SU(2)$ we cannot gain one more
from $SO(5)$ unless we incorporate yet 
another $U(1)$. In this way $SU(3)$
gets favored not only by having 
a $Z_3$ isomorphic center compared to 
the only $Z_2$ isomorphic center 
of the $SO(5)$ covering group $Spin(5)$, 
meaning 
a $3^2 = 9$ factor compared to the 
$2^2 = 4$ only for $SO(5)$, but the 
$SO(5)$ cannot get its $Z_2$ in play 
without one more $U(1)$. So in reality 
now $SU(3)$ gets in front by a factor 
$ 9$ (before one takes the $dim(G)$'th 
root.). 

So now let us compare the two groups 
obtained from the semisimple ones 
in (\ref{exa}) by cross multiplying 
them with a $U(1)$ and successively 
dividing appropriately a discrete group
out of the center:

We calculate the following goal quantities
\begin{eqnarray}
``C_A/C_F\hbox{replacement for}
(SU(2)\times SU(3) \times U(1))/Z_6 ''
&=&\left (6^2* (\frac{8}{3})^3
*(\frac{9}{4})^8\right )^{\frac{1}{12}}\\
&=&
2.957824511\\
  ``C_A/C_F\hbox{replacement for}
(SU(2)\times Spin(5) \times U(1)) 
/Z_2 ''
&=&\left (2^2* (\frac{8}{3})^3
*(\frac{12}{5})^{10}\right )^{\frac{1}{14}}\\
&=&2.54602555
 \end{eqnarray}  
Here $Spin(5) $ just stands for the 
covering group of $SO(5)$, the numbers
for the three involved simple Lie groups
are the $C_A/C_F$ ratios respectively
8/3, 9/4, and 12/5 for $SU(2)$, $SU(3)$
and $Spin(5)$(or just think $SO(5)$).
The numbers 36 and 4 come from the 
``charge'' ratio and are essentially the 
squares of the number of elements in the 
divided out subgroup in the cases here.  
The final root taking is of course because of the averaging with the dimension of the
full group- repectively 12 and 14 - to 
finally be divided out of the logarithm.

It might be nice to have in mind what 
the significance of e.g. the factor 
3 in the ``charge ratio'' $e_A/e_F$ due
to the $SU(3)$ contributes, namely a 
factor $9$ before the 12th root is taken.
Indeed $9^{\frac{1}{12}} = 1.200936955$. 
This means that obtaining the charge 
ratio due to the $SU(3)$ rather than 
there being no factor with $SO(5)$
($\approx Spin(5)$) - not even a new $Z_2$ 
to divide 
out when we already have done so using 
$SU(2)$- 
 we gain $20\% $ in the goal quantity.
The only 4.343315332\% advantage of the 
semi-simple SO(5) over the SU(3) when 
combined to $SU(2)\times Spin(5)$ and
$SU(2) \times SU(3)$ respectively is 
thus rather easily overshadowed by the 
effect of the $e_A/e_F$ from the SU(3),
which is of the order of 20 \% in the goal
quantity.

After the inclusion of the Abelian charge
type of ratio we found that the final 
advantage of the Standard Model 
{\em group} $S(U(2)\times U(3)) =
(U(1) \times SU(2) \times SU(3))/Z_6 $ 
compared to the group with the $SU(3)$
replaced by the competing $Spin(5)\approx 
SO(5)$
namely $U(1)\times SU(2) \times Spin(5)
/Z_2$ is $(\frac{2.957824511}{2.54602555} 
-1)*100\%  =16.174188\% $
\end{itemize}

\subsection{Some Property of Our 
Goal Quantity} 
We must have in mind the property of our 
``goal quantity'' due to its logarithmic
averaging 
that 
taking 
a repeated cross product of whatever 
group with itself necessarily leads to 
groups with the same goal quantiy as the 
one multiplied up. Thus if say the Standard
Model group wins, then at the same time 
any number of crossings of the Standard 
Model with itself will stand even and 
share the first place with the Standard 
Model alone.

We above essentially had the discussion 
that lead to the Standard Model winning 
except that we did not sufficiently 
carefully compare groups with different 
numbers of simple group factors. For 
instance the obvious and very serious 
competitor to the Standard Model is 
simply $U(2) = U(1)\times SU(2)/Z_2$,
which obtains the goal quantiy
\begin{equation}
``C_A/C_F\hbox{replacement for} U(2)''
= \left ( 2^2*(\frac{8}{3})^3\right )
^{\frac{1}{4}} = 2.951151786. 
\end{equation}   
This is truly exceedingly close run to
the Standard Model, but the Standard 
Model indeed wins over even $U(2)$ on the 
fourth cipher. Indeed the advange of 
the Standard Model group over the so 
closely competing $U(2)$ (which would 
physically be that there were no strong 
interactions, but only the Weinberg Salam
Glashow model say) is by
$(\frac{2.957824511 }{2.951151786}-1)
*100\% = 
.2261058 \%
$
 The contribution from the Abelian
invariant subgroup $U(1)$ namely the 
``charge ratio'' is so important we might
look for the winning group by first
taking that into account.
We may therefore look for possibilities
for the group with simple group factors 
with one combination of center groups
at a time. E.g. we could among the 
simple group combinations with one having 
$Z_2$ and one having $Z_3$, say that the
$SU(2)=A_1$ will be best to use among the 
ones with $Z_2$, while $SU(3)=A_2$ will
be ``best'' among the ones with $Z_3$.
    
We namely notice that for the same center of 
a Lie group with a simple Lie algebra
different such simple Lie algebras will
play the same role w.r.t. the division
out of the center and the ``charges for 
the Abelian group(s). 

One may rather easily see that involving 
the more complicated center groups in the 
simple Lie algebras hardly shall pay.

If you seek as would be best a $Z_k$ 
center with $k$ being prime w.r.t. the 
other $k$ values say 2 and 3 we get up to
k=5 and $SU(5)$ already has the high 
dimension 24 
and would largely weight out the 
effect of even a factor $5^2$; in 
fact $(5^2)^{1/24} =1.143529836 $.

Thus we may look at the series
and expect that the winner must be there:
\begin{eqnarray}
``C_A/C_F\hbox{replacement for} U(1)'' = \left (1\right )
^{1} = 1\\
``C_A/C_F\hbox{replacement for} (U(1)\times
SU(2))/Z_2''& =& \left (2^2*(8/3)^3\right )
^{\frac{1}{4}}\\
& =& 2.951151786 \\    
  ``C_A/C_F\hbox{replacement for} 
(U(1)\times
SU(2)\times SU(3))/Z_6''& = &\left (6^2*(8/3)^3*(9/4)^8\right )
^{\frac{1}{12}}\\
& =& 2.957824511\\
``C_A/C_F\hbox{replacement for} 
(U(1)\times
SU(2)\times SU(3)\times SU(5))/Z_{30}'' &
=&\\
 \left (6^2*(8/3)^3*(9/4)^8
*(25/12)^{24}\right )
^{\frac{1}{36}}& =& 2.341513375\\
\end{eqnarray}

We see that in this series of the most 
promising candidates with 
given centers of the covering groups for 
the simple Lie algebras
the Standard Model lies at the (flat) 
maximum.

The reader can himself check in detail and get help by studying 
our ealier work\cite{seeking}, and see that indeed the Standard Model 
wins our game with its value  2.957824511, sharply followed by the
group $U(2)$, which achieves 2.951151786 (for its silver medail).

We think it is remarkable that such a relatively simple proposal for 
a goal quantity as our slightly ad hoc 
extended essentailly Dynkin index 
$C_A/C_F$, the precise definition of 
which is largely determined from 
requirements of not depending too much 
on the notation choice, leads 
to just the gauge group that Nature has chosen! One should think that there 
is truly something in it. By this 
statement that it is largely fixed by 
independence 
of notation, we mean that it had to be a {\em ratio} of quadratic Casimirs,
if it shall be given by quadratic Casimirs at all; otherwise it would depend 
on the normalization of the quadratic Casimir, which would make it much 
more complicated to define. Our just called ``ad hoc'' extension to the 
inclusion of $U(1)$ cross product factors is really very analogous to 
the $C_A/C_F$, so indeed it is not such a series arbitrary choice.

We even have a speculative physical mechanism behind, which though might 
be thought later to be replaced by some other version.


\section{Competition Among 
Lorentz Groups
on $C_A/C_F$ and the Like}
\label{Lorentz}

The main {\em new} point of the present article 
is to present the explanation for 
why we have just 3+1 (or say just $d=4$)
space time dimensions. This explanation
is that treating the Lorentz or better 
Poincare group as ``the gauge group for 
say general relativity'' and using the 
a bit ad hoc proceedures to be suggested 
in 
subsection \ref{dgq} the 
{\em experimentally realized $d=4$ gets
singled out as having the largest ``goal
quantity'' for the ``gauge group''}. 
Here this ``goal quanity'' is taken to be 
the same one as the one that singled out 
- see section \ref{review} or 
Bennett and me 
\cite{seeking} - the Standard Model {\em 
group} by requiring it to be maximal.


\subsection{Development of Goal 
Quatities} \label{dgq}
We shall though slightly develop
the goal quantity used above for 
getting the Standard Model gauge 
{\em group} singled out, because we have
to 1) choose which group should be 
considered the ``gauge group'' relevant 
for general relativity on which to 
apply our previos game, 2) the Poincare 
group which is the best suggestion is 
for our purpose slightly unpleasant 
because it does not have nice compact 
representations of finite dimension 
like e.g. the Standard Model group had.
  
Indeed we  seek to get a statement that 
the 
experimental number of dimensions 
just maximizes some quantity, that is a 
relativly simple function of the 
group structure of say the Lorentz group, 
and  which we then call
a ``goal quantity''.

Let me therefore list some of the 
first approximation simplified 
proposals which we suggest for this 
{\em goal quantity}. But this is for the 
dimension a two step procedure: 1) we 
first use the proposals in our 
previuos article \cite{seeking} to give 
a number - a goal quantity -
for any Lie group. 2) we have to specify 
on which group we shall take and use 
the procedure of previous work;shall it 
be  the Lorentz group, its covering group 
or somehow an attempt with the Poincare 
group ? Here a series of four successive
proposals :

\begin{itemize}
\item{a.} Just take the Lorentz group 
and calculate for that the Dynkin 
index\cite{Dynkinindex} 
or rather the quantity which we already 
used as goal quantity in the previous 
article
\cite{seeking} $C_A/C_F$.
This gets especially simple for the 
except for dimension $d=2$ or smaller
semi-simple Lorentz groups
(simple in the mathematical 
sense 
of not having any invariant nontrivial
subgroup; semi-simple: no abelian 
invariant subgroup);since though the 
global structure of the Lorentz group 
is not fixed untill we assign it a 
meaning we are really here having in mind:
The Lorentz group shall ``have simple Lie 
algebra'' to apply the Dynkin index 
related ratio $C_A/C_F$ without 
further specifications. For simple groups
we can namely ignore the minor 
corrections
invented for the improvement in the case
of an Abelian component present in the 
potential gauge group.

\item{b.} We supplement in a somewhat 
ad hoc way the above {\em a.}, i.e. 
$C_A/C_F$ by taking its 
$\frac{d+1}{d-1}$th power. The idea 
behind this proposal 
is that we think of the Poincare group 
instead of as under {\em a.} only on the
Lorentz group part, though still in 
a crude way. This means we think of 
a group, which is the Poincare group, 
except
that we for simplicity ignore that the 
translation generators do not commute with
the Lorentz group part. Then we assign
in accordance with the ad hoc rule used
in \cite{seeking} the Abelian 
sub-Lie-algebra a formal replacement $1$ 
for the ratio of the quadratic Casimirs 
$C_A/C_f$: I.e. we put  
$``C_A/C_F|_{Abelean \ formal}''
= 1$. Next we construct an ``average''
averaged {\em in a logarithmic way} 
(meaning 
that we average the logathms and then 
exponentiate again) weighted with the 
dimension of the Lie groups over all the 
dimensions of the Poincare Lie group.
Since the dimension of the Lorentz group 
for $d$ dimensional space-time is 
$\frac{d(d-1)}{2}$ while the Poincare group
has dimension$\frac{d(d-1)}{2} +d = 
\frac{d(d+1)}{2}$ the logarithmic 
averageing means that we get
\begin{equation}
exp( \frac{\frac{d(d-1)}{2} 
\ln(C_A/C_F)|_{Lorentz} + 
\ln(1)*d}{d(d+1)/2})
= (C_A/C_F)|_{Lorentz}^{\frac{d(d-1)}{2}
/ \frac{d(d+1)}{2}} = (C_A/C_F)|_{Lorentz}
^{\frac{d-1}{d+1}}\label{f2} 
\end{equation}

That is to say we shall make a 
certain ad hoc partial inclusion 
of the Abelian dimensions in the 
Poincare groups. 

To be concrete we here propose to 
say crudely: Let the poincare 
group have of course d ``abelian'' 
genrators or dimensions. Let the 
dimension of the Lorentz group 
be $d_{Lor} = d(d-1)/2$; then the 
total dimension of the Poincare group 
is $d_{Poi} = d + d_{Lor} = d(d+1)/2$.
If we crudely followed the idea 
of weighting proposed in the previous 
article \cite{seeking} as if the 
$d$ ``abelian'' generators were 
just simple cross product factors 
- and not as they really are:
not quite usual by not commuting 
with the Lorentz generators - then 
since we formally are from this 
previous article suggested to 
use the {\em as if number 1 for the 
abelian 
groups}, we should use the quantity
\begin{equation}
(C_A/C_F)|_{Lor}^{\frac{d_{Lor}}{d_{Poi}}}
= (C_A/C_F)|_{Lor}^{\frac{d-1}{d+1}}
\end{equation}  
 as goal quantity. 

\item{c.} We could improve the above 
proposals for goal quantities {\em a.}
or {\em b.} by including into the 
quadratic Casimir $C_A$ for the adjoint 
representation also contributions from 
the translation generating  generators,
so as to define a quadratic Casimir for 
the whole Poincare group. This would 
mean that we for calculating our goal
quantity would do as above but 
\begin{equation}
{\bf Replace:} C_A \rightarrow C_A +C_V,
\end{equation}
where $C_V$ is the vector representation 
quadratic Casimir, meaning the 
representation under which the translation
generators transform under the Lorentz 
group. Since in the below table we in 
the lines denoted ``no fermions'' have
taken the ``small representation'' $F$
to be this vector representation $V$,
this replacement means, that we replace the
goal quantity ratio $C_A/C_F$ like this:
\begin{eqnarray}
\hbox{{\bf  (S)O(d),}}& 
\hbox{{\bf ``no spinors'':}}&\\
C_A/C_F= C_A/C_V &\rightarrow & 
(C_A+C_V)/C_F = C_A/C_F + 1\\
\hbox{{\bf Spin(d),}}&
\hbox{{{\bf ``with spinors''}}:}&\\
C_A/C_F&\rightarrow & (C_A+C_V)/C_F\\
&& =
C_A/C_F + (C_A/C_V)^{-1}(C_A/C_F)\\
&& = 
(1 + 
(C_A/C_F)|_{\hbox{no spinors}}^{-1})C_A/C_F.   
\end{eqnarray}

{\em Let me stress though that this 
proposal c. is not quite ``fair'' in as 
far as it is based on the Poincare group,
while the representations considered 
are \underline{not} faithfull w.r.t. 
to the whole Poincare group, but only 
w.r.t. the Lorentz group} 


\item{d.} To make the proposal {\em c.}
a bit more ``fair'' we should at least 
say: Since we in {\em c.} considered a 
representation which were only faithfull
w.r.t. the Lorentz subgroup of the 
Poincare group we should at least correct
the quadratic Casimir - expected crudely
to be ``proportional'' to the number of 
dimensions of the (Lie)group - by a factor
$\frac{d+1}{d-1}$ being the ratio of the 
dimension of the 
Poincare (Lie)group, $d + d(d-1)/2$ 
to that of actually faithfully represented
Lorentz group $d(d-1)/2$. That is to say
we should before forming the ratio of the
improved $C_A$ meaning $C_A+C_V$ (as 
calculated under {\em c.}) to $ C_F$ 
replace this $C_F$ by 
$\frac{d+1}{d-1}*C_F$, i.e. we perform 
the replacement:
\begin{equation}
C_F \rightarrow C_F*\frac{d(d-1)/2 +d}{
d(d-2)/2} = C_F*\frac{d+1}{d-1}.
\end{equation}        
Inserted into $(C_A+C_V)/C_F$ from {\em c.}
we obtain for the in this way made more
``fair'' approximate ``goal quantity''
\begin{eqnarray}
\hbox{``goal quantity''}|_
{\hbox{no spinor}} &=&(C_A/C_F +1)*
\frac{d-1}{d+1}\\
\hbox{``goal quantity''}|_
{\hbox{w. spinor}} &=&( 1+ 
(C_A/C_F)|_{\hbox{no spinor}}^{-1})*C_A/C_F*
\frac{d-1}{d+1}
\end{eqnarray}
{\em This proposal {\em d.} should then 
at least be crudely balanced with respect
to how many dimensions that are 
represented faithfully.}  

\end{itemize}


\subsection{Calculation of Goal 
Quantities}
Let us now begin 
listing the 
values of these ``goal quantities'' for 
the 
Lorentz groups for the various numbers $d$
which the dimension of space time might 
take on.

In the first table we give  
the  ``goal quantity''
{\em a.} and the in order to 
go crudely towards the Poincare group
``goal quantity'' {\em b.}.

{\bf Caption:} In the below tables
we evaluate for different dimensions $d$
of the Minkowski space-time - for 
simplicity here replaced by the Euklideanized
d dimensional space-time, but that makes
no difference for our calculation here -
the first two goal-quantities proposed,
{\em a.} and {\em b.} in subsection 
\ref{dgq} written in 
respectively 5th and 7th columns.
Because of the ambiguity of the global
structure of the Lorentz-{\bf group}
the group in $d$ dimension may be either
O(d) (essentially SO(d) if we do not 
include parity) or Spinor(d) if we use the covering 
group. We have 
therefore for each value of the dimension
$d$ two items corresponding to these 
two global extensions of the 
Lie algebra of the Lorentz group, and 
they are denoted by ``no spinors'' and 
``with spinors'' respectively.

\begin{center}
    \begin{tabular}{ | l | l | l | l 
|l|l|l|}
\hline
Dimension & Lorentz group & Spinor or not
& Rank& Ratio $C_A/C_F$& 
$\frac{d_{Lor}}{d_{Poi}}$& 
$(C_A/C_F)^{\frac{d_{Lor}}{d_{Poi}}}$ \\
  &  &   & &max a)& &max b)\\
d=1 & discrete &&0 & -& 0& indefinite $0^0$ \\
\hline
d=2\footnote{Because of the Lorentz group being Abelian our treatment of 
dimension d=2 can only be formal; use $C_A/C_F =1$ for 
abelian and correct for dividing out subgroup of the center like in
\cite{seeking}} & (S)O(2) = U(1)&no spinor&1 & - (formally 1)
&1/3& - (formally 1)\\
&U(1)& with spinor&1& -(formally 2)&1/3&-(formally $2^{1/3}=1.26$)\\
\hline
d=3 & (S)O(3) & no spinor&1& 1&1/2&1\\

   & Spin(3)=SU(2)& with spinor&1& 8/3=
2.6667 &1/2 &$\sqrt{8/3}$
\\
&&&&&&=1.632993162\\
\hline
d=4 & (S)O(4) & no spinor &2& 4/3=1.3333& 3/5&$ (4/3)^{3/5}$
\\
&&&&&&  =1.188401639\\
   & Spin(4) 
&with 
spin&2&
8/3=2.6667& 3/5 & $(8/3)^{3/5}$
\\
& = $SU(2)\times SU(2)$ &&&&&=1.801280051
 \\
\hline
d=5 & (S)O(5)& no spinor&2& 3/2=1.5& 2/3& 
$(3/2)^{2/3}$
\\
&&&&&& =1.310370697\\
   & Spin(5) & with spinor&2& 12/5=2.4&2/3&$ (12/5)^{2/3}$
\\
&&&&&&=1.792561899\\
\hline
d=6 & (S)O(6) & no spinor &3& 8/5=1.6&5/7&
$(8/5)^{5/7}$ 
\\
&&&&&&= 1.398942897\\
   &Spin(6) = SU(4)& with spinor&3& 
32/15=2.1333
& 5/7 & $(32/15)^{5/7}$ 
\\
&&&&&&=1.718074304\\

\hline
d=7 & (S)O(7)& no spinor &3& 13/7=1.8571
& 3/4&
$(13/7)^{3/4}$ 
\\
&&&&&&=1.590867407\\
   & Spin(7)& with spinor&3& 40/21=1.9048 & 3/4&
$(40/21)^{3/4}$ 
\\
&&&&&&=1.621363987\\
\hline
d=8 & (S)O(8)& no spinor &4& 12/7=1.7143 & 7/9 & $(12/7)^{7/9}$ 
\\
&&&&&&=1.520774129\\
   & Spin(8) & with spinor&4& 12/7=1.7143 & 7/9&
$(12/7)^{7/9}$ 
\\
&&&&&&=1.520774129\\
\hline
d=9& (S)O(9) & no spinor&4& 7/4=1.75 & 4/5 & 
$(7/4)^{4/5}$ 
\\
&&&&&&=1.564697681\\
  & Spin(9) & with spinor&4&14/9=1.5556 & 4/5& 
$(14/9)^{4/5}$ 
 \\
&&&&&& = 1.423994858\\
\hline
d=10& (S)O(10) & no spinor&5& 16/9=1.7778 & 9/11 &
$(16/9)^{9/11}$ 
\\
&&&&&&=1.601198613 \\
   & Spin(10) &with spinor&5&64/45=1.4222 & 9/11&
$(64/45)^{9/11}$
\\
&&&&&&=1.33399805\\
\hline
d=11 & (S)O(11)& no spinor&5& 9/5=1.8 & 5/6 & 
$(9/5)^{5/6}$ 
\\
&&&&&&=1.632026054\\
   & Spin(11) & with spinor&5& 72/55=
1.3091 & 
5/6 & $(72/55)^{5/6}$ 
\\
&&&&&&=1.251626758\\
\hline
d=12 & (S)O(12) & no spinor&6& 44/23=
1.9130 & 
11/13& $(44/23)^{11/13}$ 
\\
&&&&&&=1.731340775\\
    & Spin(12) & with spinor&6&40/33=
1.2121 & 
11/13& $(40/33)^{11/13}$
\\
&&&&&&=1.176773318\\
\hline
d=13& (S)O(13) & no spinor&6& 25/13
=1.9231 & 6/7& 
$(25/13)^{6/7}$ 
\\
&&&&&&1.75156277\\
   & Spin(13) & with spinor&6&44/39
=1.1282 &6/7&
$(44/39)^{6/7}$ 
\\
&&&&&&=1.108929813\\
\hline
d=14 & (S)O(14)& no spinor&7& 24/13
=1.8461 & 
13/15& $(24/13)^{13/15}$ 
\\
&&&&&&= 1.701239682\\
   & Spin(14)& with spinor&7&104/105
=0.9905 & 
13/15& $(104/105)^{13/15}$
\\
&&&&&&=0.991740772\\
\hline
\hline
\end{tabular}
\end{center}

\begin{center}
 \begin{tabular}{ | l | l | l | l 
|l|l|l|}
\hline
Dimension & Lorentz group & Spinor or not
& Rank& Ratio $C_A/C_F$& 
$\frac{d_{Lor}}{d_{Poi}}$& 
$(C_A/C_F)^{\frac{d_{Lor}}{d_{Poi}}}$ \\
  &  &   & &max a)& &max b)\\
d odd& (S)O(d) & no spinor&n =(d-1)/2& 
2- 1/n 
&$ \frac{d-1}{d+1}$& $(2-\frac{1}{d-1})^{\frac{d-1}{d+1}}$ \\
& & && =2 -2(d-1)& & \\
    & Spin(d) & with spinor&n=(d-1)/2&
$\frac{8(2n-1)}{n(2n+1)}$ 
&$\frac{d-1}{d+1}$&$(\frac{16(d-2)}
{d(d-1)})
^{\frac{d-1}{d+1}}$  \\
&&&&  =$\frac{16(d-2)}
{d(d-1)}$ && \\
\hline
d even& (S)O(d) & no spin& n=d/2&
$\frac{4(n-1)}{2n-1}$ 
&$\frac{d-1}{d+1}$&$(\frac{2(d-1)}{d-1})
^{\frac{d-1}{d+1}}$\\
&&&&=$ \frac{2(d -2)}{d-1}$&&\\
   & Spin(d)& with spinor& n=d/2&
$ \frac{16(
d-2)}{d(d-1)}$&$\frac{d-1}{d+1}$&$ (\frac{16
(d-2)}{d(d-1)})^{\frac{d-1}{d+1}}$\\ 
\hline
d $\rightarrow \infty$ &    & no spinor&
c$\rightarrow \infty$& $\rightarrow$ 2&
$\rightarrow$ 1&$\rightarrow $ 2\\
   & & with spinor& $\rightarrow \infty$&
$\rightarrow$ 0&$\rightarrow$ 1& $\rightarrow $ 0\\
\hline
\hline

\end{tabular} 
\end{center}

${}^1$ The case d=2 is special because 
the Lorentz group is Abelian $U(1)$ for
d=2, and we must apply the formal 
extension definition $C_A/C_F =1$
for $U(1)$ from our previous work 
\cite{seeking}, and even include an
extra factor connected with dividing out
a subgroup of the center, or better say
that the formal quadratic Casimir shall
behave like the charge squared for the 
$U(1)$.

\subsection{Discussion of Table}

Motivated either by the fact that we have 
spinor transforming particles in nature 
- namely the Fermions - or because the 
goal-numbers for the spinor groups 
are anyway the biggest (most competitive)
we should think of
the Lorentz {\em group} as the spinor 
group and therefore in the above table 
read the Spin(d) entrances rather than 
the  $(S)O(d)$-entrances:  

Concentrating on the $Spin(d)$-entrances
we then find that with the proposal 
{\em a.} of subsection \ref{dgq} the 
dimensions  d = 3 
and d= 4 stand even with the same 
goal-number 8/3 = 2.6667. But note that 
at least the experimental dimension 4 
already is in the sample of the 
``winners'' with the simple choice of 
{\em a.} meaning, that we only consider 
the genuine Lorentz group - while totally
ignoring the Abelian part of the Poincare 
group. 

Next when we go to the slightly 
more complicated version of a 
goal-quantity, namely {\em b.}, we get 
the separation between also d=3 and d=4,
and it is the d=4 dimension, that ``wins'',
because we get for d=3 only 1.6330, while 
we for d=4 we obtain 1.8013. Thus in this 
approximate treatment of the Abelian part 
also being included the (after all rather)
``little'' difference between the two 
schemes {\em a.} and {\em b.} leads to 
giving the d=4 case - the experimental
case  - the little push forward making 
the experimental dimension d=4 be the only 
winner!

{\bf Caption:}
We have  put the goal-numbers for the 
third proposal {\em c} in which I (a bit 
more in detail) seek to make an analogon to 
the number used in the reference 
\cite{seeking} in which we studied the
gauge group of the Standard Model.
The purpose of {\em c.} is to approximate 
using  the {\em Poincare 
group} a bit more detailed, but still 
not by making a true representation 
of the Poincare group. I.e. it is still 
not truly the Poincare group we represent 
faithfully, but only the Lorentz group,
or here in the table only the covering 
group $Spin(d)$ of the Lorentz group.
However, I include in the column 
marked ``{\em c.}, max c)'' in the 
quadratic Casimir $C_A$ of the Lorentz 
group an extra term coming from the 
structure constants describing the 
non-commutativity of the Lorentz group 
generators with the translation generators
$C_V$ so as to replace $C_A$ in the 
starting expression of ours $C_A/C_F$
by $C_A + C_V$. In the column marked 
``{\em d.}, max d) '' we correct the ratio
to be more ``fair'' by counting at least 
that because of truly faithfully 
represented part of the Poincare group 
in the representations, I use, has only 
dimension $d(d-1)/2$ (it is namely 
only the Lorentz group) while the full
Poincare group - which were already in 
{\em c.} but also in {\em d.} used in 
the improved $C_A$ being $C_A +C_V$ -
is $d(d-1)/2 + d = d(d+1)/2$. The 
correction is crudely  made by the 
dimension ratio $dim(Lorentz)/dim(Poincare)
=(d-1)/(d+1)$ given in the next to last
column.  

\begin{center}
 \begin{tabular}{ | l | l | l | l 
|l|l|l|}
\hline
Dimension & Lorentz group 
& Ratio $C_A/C_F$ &Ratio $C_A/C_V$& 
{\em c.}-quantity& $\frac{d-1}{d+1}$
& {\em d.}-quantity \\
  &(covering group)  &for spinor   
&as ``no spinor'' 
&max c)& &max d)\\
\hline
2\footnote{the treatment of the case d=2 becomes because of the 
Abelianness of Lorentz group illdefined, and we have to use the somewhat formal specification from \cite{seeking}; for the ``with spinor'' we even use 
formally the special rule connected with dividing out a subgroup of the center,
but since d=2 gets beaten by the higher dimensions the details are not so serious for us.}  & U(1)& -(formally 2)& -(formally 1)&4&1/3&4/3=1.33\\ 
\hline
3& spin(3) & $\frac{8}{3} = 2.6667$&1&
$\frac{16}{3}=5.3333$ 
&$ \frac{2}{4}=.5$& 
$\frac{8}{3}= 2.6667$ \\
\hline
4&$Spin(4)$  &$\frac{8}
{3}= 2.6667$ &$\frac{4}{3}$& $\frac{14}{3}
 =4.6667$ &$\frac{3}{5}=.6 $ &$\frac{14}{5}
= 2.8$ \\
&$=SU(2)\times SU(2)$&&&&&\\
\hline
 5   & Spin(5) &$\frac{12}{5}=2.4 $&$\frac
{3}{2} =1.5$&
$4$  
&$\frac{4}{6}=.667$&$\frac{8}{3}= 2.6667$  
\\
\hline
6 &$Spin(6)$&$\frac{32}{15}$&$\frac{8}{5}
=1.6$
&$\frac{52}{15}=3.4667 $  &
$\frac{5}{7}=.714$&
$\frac{52}{21} =2.4762$  \\
\hline
d odd & Spin(d)& 
$\frac{8(2n-1)}{n(2n+1)}$&$2- 1/n$   &
$\frac{8(3d-5)}{d(d-1)}$&$\frac{d-1}{d+1}$&
$\frac{8(3d-5)}{d(d+1)}$\\
&&$=\frac{16(d-2)}{d(d-1)}$&$=2-2/(d-1)$ &&&\\
\hline
d even& $Spin(d)$ & $ \frac{16(
d-2)}{d(d-1)}$&$\frac{4(n-1)}{2n-1}=\frac{
2d-4}{d-1}$ &
$\frac{8(3d-5)}{d(d-1)}$ &
$\frac{d-1}{d+1}$
&$\frac{8(3d-5)}{d(d+1)}$ \\ 
\hline
d odd $\rightarrow \infty$ & Spin(d) & 
 $\approx 16/d$ &$\rightarrow 2$    
& $\approx 24/d$&$\rightarrow 1$& 
$\approx 24/d \rightarrow 0$\\
\hline
d even $\rightarrow \infty$ & Spin(d) & 
 $\approx 16/d$ &$\rightarrow 2$    
& $\approx 24/d$&$\rightarrow 1$& 
$\approx 
24/d \rightarrow 0$\\
\hline
\hline

\end{tabular} 
\end{center}
${}^2$ The case d=2 is only getting its
$C_A/C_F$ rather formally by seeking to
use the rules roughly of our previous 
article\cite{seeking}, because the Lorentz
group is Abelian.

We see from the table here, for simplicity 
made only for the most  competitive  case 
of ``with spinor'' in the terminology 
the foregoing table, that with column 
{\em c.}-goal numbers actually it is 
d=3 rather than the experimental dimension
d=4 that ``wins''(i.e. is largest), but 
this series of proposed numbers {\em c.}
is not truly ``fair'' in as far as one has 
effectively used only the Lorentz group 
in the denominator $C_F$ but at least 
crudely the full Poincare group in the 
numerator $C_A+C_V$. Thus in order to 
avoid a simple wrong expected variation 
of a quadratic Casimir with the 
dimensionality of the Lie group, we 
should at least correct the denominator 
$C_F$ by multiplying it by the ratio
of the dimension of the Poincare Lie  
group over  that of the Lorentz Lie group,
$(d+1)/(d-1)$. When we make this 
``fairness correction'' at least crudely 
getting no obvious wrong 
Lie-group-dimension-dependent factor
in, then the dimension d=4 becomes 
(again) the winner! In fact we get for 
d=4 (the experimental dimension) 
the goal quantity in column {\em d.} 
equal to 14/5= 2.8 while  accidentally
the two neighboring dimensions d=3 and d=5 both gets in stead 8/3 = 2.66667, 
which 
is less. 

But notice that it is a rather 
smoothly peaked curve with the peak near the 
experimental dimesnion 4, so that the 
latter becomes the winner among integers,
but it is only by a tiny bit it wins. That 
is to be expected from the smoothness of 
the variation of the goal number with 
the dimension d. This smallness of the 
excess making the d=4 be the winner of 
course makes the uncertainty bigger and my 
``crude '' corrections rather than exactly 
calculating some welldefined quantity 
is thus not so convincing. Anyway I think, 
that the accuracy may be good enoungh, and 
the simplicity of the proposed goal 
quantities sufficient to make it at 
least highly suggestive, that the coincidence of
 the winning dimension and the 
experimental one means that we are on the 
right track!

\section{Conclusion}
\label{conclusion}
We have found that a couple of very 
reasonable specifications of what the 
extension of our previous \cite{seeking} 
quantity should be  to 
be maximized to obtain the Standard Model
gauge group leads to that the maximization
of the generalized quantity gives as the 
``winner'' the dimension d=4 as is 
empirically  the dimension! That is to
say we have found a possible 
{\em explanation} for, why we have just 
4 (meaning 3+1) dimensions of space-time!
  
In fact we have extended the main idea 
of claiming the {\em maximization} of 
essentially the to the Dynkin index 
related group dependent quantity 
$C_A/C_F$ (with $C_A$ and $C_F$ being 
the quadratic Casimirs for respectively
the adjoint representation $C_A$ and for 
that (essentially) faithful representation
$F$ chosen so as to maximize the ratio
$C_A/C_F$.) to lead to the experimentally 
realized group (the Standard Model group). 
For honesty it should be admitted that the 
victory of exactly the Standard Model Group
were dependent on our slightly ad hoc treatment 
of the Abelian invariant subgroup - i.e. 
the $U(1)$ - needed because of the ratio 
$C_A/C_F$ being rather meaningless a 
priori 
for an abelian group. For honesty we must also admit 
that in reference \cite{seeking} we 
sneaked in 
a dependence on the {\em group} rather 
than 
only {\em Lie algebra} by considering the 
volume of the {\em group} which depends 
on the 
identification of center elements, properties 
being revealed phenomenolgically via the representations
of the gauge group (except for the case of d=2 this
detail is though not relevant for the Lorentz group and thus the 
dimension). In the review \ref{review}
we saw that the Standard Model group
came quite remarkably out as the group 
with the highest goal quanitity! This 
should be considered a very remarkable 
victory for our type of scheme because 
there are a lot of groups which a priori
could have been the gauge group relevant 
for nature.   

Our extension consists in that we in 
stead of 
the gauge group  rather consider the 
{\em Lorentz group}, or we even seek to 
use 
the {\em Poincare group}. Then of course 
our quantity $C_A/C_F$ or slight 
modifications/``improvements'' of it 
- enumerated  {\em a., b., c., d.} - 
will depend on the dimension d of 
space time. The dimension $d$   gives of course a
different  
Lorentz group for each value of $d$. 
We then inserted this $d$-dependent Lorentz group
in stead of the gauge group which were studied 
in last paper\cite{seeking}. The various modifications, 
  {\em a., b., c., d.}, shall be considered attempts 
to use the {\em Poincare group} instead of the 
Lorentz group, but rather than truly doing that,
make some approximate treatment as if  crudely using 
 the Poincare group. 

The results of the search for 
the dimension having  the largest 
``goal-quantity'', using various 
proposals for the exact form such 
as {\em a.}
meaning  
$C_F/C_F$ simply, are the following:

\begin{itemize}
\item{a.} The simple quantity $C_A/C_F$
for the {\em Lorentz group} with the same
formal assignment for abelian group as 
used in \cite{seeking}, here making d=2
noncompetitive(but at least having a 
score formally),
  leads to d=3 and d=4 
standing even, both scoring the same 
number $C_A/C_F = 8/3$.

\item{ b.} Making a crude correction 
to consider instead the quantity 
$(C_A/C_F)^{(d-1)/(d+1)}$ leads to that 
the experimental dimension of space time 
d=4 gets the largest score. The meaning 
of this slight modification of {\em a.}
is that we make an attempt to take the 
group to replace the gauge group in our previous paper\cite{seeking}  
to be the Poincare group rather than the 
Lorentz group. We, however,
only make a crude attempt in that 
direction.
Since the Poincare has the translation 
subgroups, 
which are by themselves Abelian we 
naturally tend to use 
formally - just like in 
reference \cite{seeking} - to assign a 
factor
1 to the Abelian groups.  Then we  average in the logarithm our goal numbers for 
the various 
factors into which the group falls 
weighting with the dimension
in the Lie algebra.

  The inclusion of the 
Poincare group is not done in a fully 
correct way though 
 in as 
far as we only consider the faithful
representations of the Lorentz group
and only extend a bit speculatively to 
weight as if we had the Poincare group.

\item{ c.} Still thinking of crudely
using the Poincare group rather than 
the Lorentz group, we proposed 
to still take a representation $F$ only 
of the Lorentz group, but evaluating 
the  quadratic Casimir for the Poincare
group, although that sounds not quite 
``fair''. The quadratic Casimir we 
used here under {\em c.} for ``the 
Poincare group'' were taken to $C_A+C_V$,
where $V$ denotes the vector 
representation and thus $C_V$ its
quadratic Casimir. In this ``unfair''
game the dimension for space time d=3=2+1 
got the highest score. So our hoped for 
victory of the experimental dimension 
failed in this ``unfair'' proposal.
{\em But since I stress the ``unfairness''
of this proposal, we should not take 
this proposal seriously.}

\item{d.} This last proposal in the 
present 
article is a crude attempt  at least to 
correct for, that the ratio of the 
dimensions of the Poincare and the Lorentz
Lie groups is space-time dimension $d$ 
dependent. That is to say, we argue that 
the quadratic Casimir $C_F$ for the 
representation $F$ of the Lorentz group
should at least be scaled so as to 
correspond to a representaion rather 
of the Poincare group by being multiplied
by the ratio of the Lie group dimensions 
of the Poincare group relative to that 
of the Lorentz group, $\frac{d(d-1)/2+d}{
d(d-1)/2} = \frac{d+1}{d-1}$. That is to 
say we perform the crude correction of
replacing
\begin{equation}
C_F \rightarrow C_F*\frac{d+1}{d-1}.
\end{equation}       
Since the quantity $C_F$ occurs in the
denomiator of the quantity $(C_F+C_V)/C_F$
maximized under {\em c.} of course this
quantity is scaled the opposite way, and 
the goal-quantity in this proposal 
{\em d.} is taken as
\begin{equation}
\hbox{``goal quantity {\em d.}''}
= \frac{(C_A +C_V)(d-1)}{C_F(d+1)}.
\end{equation}
Now the result becomes that {\em the 
experimental dimension d=4 has the 
largest value for the goal quantity d.},
in as far as it gets
\begin{equation}  
\frac{(C_A +C_V)(d-1)}{C_F(d+1)}|_{d=4} =
\frac{14}{5} = 2.8,
  \end{equation}
while by accident the two neighboring 
space time dimensions 3 and 5 score only 
$\frac{8}{3} = 2.6667$. 
So indeed the {\em experimental space time 
dimension 4 won the most developped 
suggestion d.}. 
\end{itemize}

This means that apart from the ``unfair''
proposal c., all the four  proposals
here have the space time dimension
d=4 realized in nature obtain a largest
``goal quantity'' among the winners!
In a. d=3 and d=4 share the winner place,
but in the two other ``fair'' proposals
{\em b.} and {\em d.} it is indeed space
time dimension d=4, the experimental one, 
that gives the highest ``goal quantity''.

Taking this result serious, and not as 
being just accidental coincidence or
a result of inventive construction, 
it must tell us about the reason for
that the dimension became just d=4.
Taking our result serious we must look 
for what is the spirit behind the 
proposals above, so as to obtain an 
answer to ``Why did we get just d=4 
space time dimensions?''. This ``spirit''
behind the proposals here set up to 
select the experimental dimension d=4
seems to be that the group - the 
Poincare or Lorentz group or the 
gauge group say - should be {\em 
representable in a way where the 
matrices or other objects on which 
the group is represented are relatively
slowly varying under the group}. 
 
We may, if taking this ``slowness''
of the motion of the representative 
in the representation with the group
element serious, seek to invent a model
behind the 4-dimensional space time and 
the Standard Model gauge group that could 
explain that slowness. One possible such 
explanation could be:

Truly the fundamental physics model or
theory is ``random'' and that  without the 
symmetries we seek to explain. Then 
``by accident'' there appears 
approximately some 
symmetries - and we here hope for say 
some Lorentz invariance symmetry -.
Now we dream that there may be some way 
in which such an approximate symmetry can 
be automatically become exact in practice.
We, F{\o}rster, Ninomiya and 
me \cite{Foerster}(see also Damgaard et al. \cite{Damgaard}),  
have 
actually argued that gauge symmetry with 
electrodynamics (and Yang Mills theory)
as example can occur in a whole phase 
in practice giving  precisely the 
{\em massless} photon in that phase. Thus
we can for symmetries that can somehow 
be considered gauge symmetry - as can 
the Lorentz symmetry in general 
relativity - speculate that such 
symmetries appear in practice as exact
providided though, that they are there 
approximately at 
first\cite{Lehto, Lehto2}.. 
But now the crux of the matter is that 
{\em if a symmetry is represented by 
slowly moving matrices or whatever, then
one must expect that statistically it 
would be easier to get the symmetry 
approximately by accident}. If it were 
such that the fundamental theory could 
be considered random and only obtaining 
some symmetries by accident - at first 
approximately, but perhaps made exact 
by some 
mechanism\cite{Foerster,Lehto, Lehto2, 
Damgaard} - 
then we could consider
the practically random Lagrange or action 
as taking random values for regions of 
some (small) size in the value space for
the {\em representaion} of the group
which gives the transformation properties
of the fields or degrees of freedom  under the group in question.
Now when a group is represented by a 
representation, which in some sense is
the represented matrix or field, and these fields or matrices  move 
slowly for an appropriately normalized 
motion of the group element represented,
then one can vary the group element much
before one varies the representation field 
much. But this means that one needs less 
good luck to get a symmetry accidentally
the slower the represention moves, because
the displacement inside the group 
(itself) corresponding to one of the 
(small) size regions (over which we 
assume essential constancy of the action)
becomes bigger the slower the 
representation motion rate.

The crucial point should be that one 
would with the in some sense random action
have a better chanse to obtain by accident
a certain symmetry, when this symmetry is 
represented on the fields or degrees of 
freedom by a ``slowly moving 
representaion'', so such a symmetry would 
more often occur by accident, if one 
thinks this random action way. 

So when our various ``goal quantities''
favour the experimentally found gauge 
group and the dimension of space time,
it means that the groups realized in 
nature are the ones that have the optimal
chanse to come out of a random action 
model. This is so because these goal 
quantities being large means that the 
representation motion is slow. 

So the message from the gauge group and
the dimension is that such a random 
action philosophy is one possible mechanism
behind the choice by nature of the gauge groups
and dimension. 

The idea of there being made in some 
sense a lot of attempts randomly of 
groups to be tested off could be said to 
have remiscense of the idea of 
a gaugeglass \cite{gaugeglas}, which 
though rather means the action is 
random quenched randomly locally, but 
that the gauge group is given from the 
start; but the spirit is similar.

A priori one should speculate about 
possible other physical machineries that 
could 
explain that precisely our type of 
``goal quantities'' should point to 
realized gauge group and dimension of 
space time; but at first it seems
that the random action type of model 
allowing symmetries of the type with 
highest goal quantities is a good idea
and very likely something like that could
be the reason behind the choice by nature
of the gauge group (of the Standard 
Model) and of the dimension. 

In any case we have found a surprisingly
simple principle - the maximization of
our to each other rather closely related 
``goal quantities'' - leading to 
both the gauge group of the Standard Model
{\em and} the dimension of space time 
being 4. 

Let me stress that a work of the present
type and of \cite{seeking} - finding a 
goal quantity leading 
to the realized groups - is an attempt 
to ask in a  phenomenological way, whether there 
is some signal in the details of the 
 presently by phenomenology supported theory that 
successively can give us hint(s) about 
the more fundamental theory behind the 
presently working Standard Model with 
its gauge group $S(U(2)\times U(3))$
and the seeming dimension 4.          
     
\section{Acknowledgement}
I would like to thank the Niels Bohr 
Institute for allowing me status as 
professor emeritus and for economical 
support and Matjaz Breskvar, support for visiting the Bled Conference
``Beyond the Standard Models'' wherein 
the first paper\cite{seeking} in the 
present series of 
two articles were presented together 
with Don Bennett - although a year earlier though -, who is also thanked 
together with colleagues discussing there
the previous work.       

\section*{Appendix, Earlier works}
\label{earlier}

\subsection{Why the Standard Model 
Group}
 The first question, which the present 
series of articles, namely this and the
previous work\cite{seeking}, attacks is
``Why did Nature choose just the Standard 
Model group $S(U(2)\times U(3))$, it could 
seemingly have chosen among a lot of 
Lie groups?'' Historically this Standard
Model group has appeared as pieces arising
from different types of interactions, 
one $U(1)$-subgroup came from 
electromagnetic interactions, built in  a 
bit complicated way into what as {\em 
group} stands as $U(2)$ 
(namely $SU(2)\times U(1)$ 
Lie-algebra-wise) and contains also the 
weak 
nuclear forces.  Finally the QCD 
describing strong interaction connected 
to the $SU(3)$ part of the Standard Gauge
group were added. The Standard model is 
so to speak
by phenomenology found piece-wise: 
sub-algebra for sub-algebra. It is first 
afterwords or at least by 
inclusion of further possible pieces 
of the gauge group that speculations 
of grand unification models of various 
types have appeared. 

The specification of some appropriate 
GUT model\cite{GUT}, say as the simplest 
and most
promising $SU(5)$, together with some 
breaking scheme, makes up an explanation 
for the Standard Model gauge Lie algebra
and also easily for the gauge {\em group}
- which we take to be implemented as 
a restriction on which representations
are allowed, so that we can indeed 
claim a by phenomenology accessible 
element of knowledge - being the one 
realized in nature. Truly a major 
part of the success of the GUT $SU(5)$
model is that the representations of 
the $SU(5)$ gauge group are automaticly 
representations of the subgroup of 
the $SU(5)$ with the Lie algebra of the 
Standard Model, and this sub{\em group}
is just the $S(U(2)\times U(3))$, so that
the GUT $SU(5)$ precisely can explain 
the same restrictions on the allowed 
representations as can the Lie{\em group}
$S(U(2)\times U(3))$. If we can explain 
this {\em group} structure and not only
the Lie algebra structure of the Standard 
Model group, then we would have less 
need for grand unification, because it 
would mean obtaining similar predictions 
for the representations of say quarks 
and leptons under the gauge group.   
It is simply so that $S(U(2)\times U(3))
\subset SU(5)$ while e.g.the group
$U(1)\times SU(2) \times SU(3)$ is 
\underline{not} a sub{\em group} of 
$SU(5)$. Therefore the restrictions
on the representation from $SU(5)$ 
interpreted as containing the Standard 
Model after some breaking leads to 
the restrictions of the subgroup 
$S(U(2)\times U(3))$. Other grand unified
groups that have success typically 
contains $SU(5)$ as a subgroup and thus
can reproduce the same representation
restrictions corresponding to the 
Standard Model \underline{group}
$S(U(2)\times U(3))$. 
Especially we can mention the 
$SO(10)$-group, which comes out of 
\cite{Norma}.  
In principle 
a different model behind the Standard 
Model is the flipped 
$SU(5)$\cite{flippedSU5}.    
        
Often super string theories would go 
through some of these grand unification 
type models, but one can also construct 
models going to the Standard Model 
without going via the unifying groups.

I think one can say that the possibilities
are so many that one should admit that 
it is only if  you somehow have already
got to know the grand unified group,
that there is much predictive power 
as to what group we get in practice.
Otherwise the resulting group could be so 
many different possible ones that 
there is not much predictive power 
in these models. 

So only if one has a prediction of 
the unified group in a model should 
we consider the standard model group 
explained by grand unification.

In this respect the unification 
of spin and charge model of 
Norma Mankoc Borstnik (et al.)\cite{Norma}
is better by leading to $SO(N)$ groups 
just from the spirit of it. To get 
just $SO(10)$ as is needed for 
getting the Standard Model the 
dimension 13 +1 should be put in, and 
so after all the prediction of the 
Standard Model does not come quite 
without seemingly ad hoc numbers 
being put in. We think of the number 14 
for the space time dimension. 

The type of explanation for the Standard
Model group, which we were after in the 
present series of works is rather to seek 
to characterize the group by properties 
defined for abstract Lie-group, and then 
postulate some number-valued 
group-characterization - which we should 
guess so cleverly that it will specify
just the wanted Standard Model group -
to be by Nature arranged to be  maximal.

One attempt of this type were our work,
Niels Brene et al. \cite{Brene} in which 
we 
define a concept of ``skewness'' in a 
quantitative way, so that one obtains 
a number for each Lie group (we never 
got the idea completely developed to 
specify how many $U(1)$ invariant 
subgroups there should be, but I think
we essentially get to there being just one
$U(1)$ factor).
If we to help the project a bit assumed 
that we should only consider the possible
gauge groups with just one $U(1)$-factor 
and construct a measure for the degree 
of symmetry as the logarithm of the 
number of outer automorphisms $o$ divided 
by the rank of the Lie algebra $r$
\begin{equation}
``symmetry'' = \frac{\ln(o)}{r},
\end{equation} 
  we get the Standard Model 
\underline{group} to have the minimal 
value of this quantity ``symmetry''.
In this sense the Standard Model group
is among the most ``skew'' in the sense 
of being the least symmetric.

The subject of the present series of 
papers \cite{seeking} and the present 
article is a different approach of 
the same character as the just mentioned 
``skewness'' characterization of the 
Standard Model group. However, our 
present attempt to characterize the 
gauge group is by a different quantity 
from the ``symmetry'' quantity; it is 
a new attempt. It must though be admitted
that there appears one overlapping element
in the two different quantities to be 
extremized: some of the potential outer 
automorphisms that are to be counted in 
$o$ can be gotten away with by dividing 
out of the group a subgroup of the center.
Actually such a division out of a subgroup
of the center seems to be a rather 
characteristic property of the Standard 
Model \underline{group} so that it is 
quite helpful if a quantity to determine 
the Standard Model group has a strong 
dependence on the appearance of an 
effect of such a division out of a 
subgroup of the center, so as to favour 
a gauge group with a lot of division out 
of the center. This division out 
is not relevant in the very first 
version of  our present proposal,
namely the ratio $C_A/C_F$ of the 
quadratic Casimir of the adjoint 
representation $A$, denoted $C_A$ 
to the quadratic Casimir of that
faithful representation $F$ having 
the smallest quadratic Casimir $C_F$.
However, when we began to help the 
construction of our quantity by specifying
the details concerning the Abelian $U(1)$ 
components in the Lie group for which 
our quadratic Casimir ratio is not a 
priori well-defined, we managed to -
one could almost say - ``sneak in''  the
division out, so that we by the details 
of defining what to do when we have 
$U(1)$ factors get the groups with a 
complicated division out of the center get 
favored to win the game and become the
gauge group chosen by nature.

In this way our present proposal for what
nature has chosen to maximize and our 
older proposal of maximizing ``skewness''
are not completely different because they 
have an overlap by both favouring the 
complicate division out of the center.
But apart from that they look 
superficially seen very different.
Thus our present proposal is quite new 
after all. 

  I shall review the successive small 
improvements in finding partly 
phenomenologically our final
suggestion for the quantity or the game 
that should specify the gauge group 
to be chosen in section \ref{review}
below.

\subsection{History of 
Explaining 3+1=4
Dimensions}

As said the main purpose of the present
article is to use the idea from our 
earlier article \cite{seeking} to explain 
why nature should have chosen just 4
(meaning 3+1) space time dimensions.
This way to be explained below is new 
relativ to earlier attempts to explain, 
why we shall just have four dimensions:

One of the earlier attempts is my own 
\cite{RD} starting the idea of ``Random 
Dynamics'' by pointing out that in a 
{\em non-Lorentz invariant } theory - 
being a quantum field theory in which 
neither rotational nor boost invariance 
is present, but only translational 
invariance- one finds genericly
that assuming an appropriate Fermi-surface
an {\em effective Lorentz invariance 
with 3+1 dimensions} appears automaticly!
In this sense I claimed to derive under 
very general assumptions - as almost 
unavoidable - the appearance of both 
Lorentz and thereby rotational invariance 
and of just the right number of dimensions,
4=3+1. The success of this dimension-post-diction \cite{RD} 
were for me the 
introduction to a long series of works 
seeking to derive from almost nothing or from
a random theory - not obeying many of the 
usual principles - which I gave the name
``Random Dynamics'' \cite{RD,Foerster, 
RDrev, Astri, RDDon1}, many of the known physical 
laws. Really we may consider 
the present work as an alternative attempt
to derive the dimension of space time, 
much in Random Dynamics way, in as far as
we ended up suggesting the philosophy that 
it would be most easy/likely to get 
just the 
experimental dimension by accident.
(If successful then of course the present work would 
be a second derivation of 3+1 dimensions in Random Dynamics). 

Also Max Tegmark has derived 3+1 
dimensions
from a similar Random Dynamics like 
philosophy of ``all mathematics being 
realized''\cite{Tegmark}.
Max Tegmark considers the differential 
equations for
the time development of fields so as to 
guarantee 
equations with predictivity, as well as 
the stability of
atoms. For the anyway from his arguments 
needed case
of just one time his figure shows that 
the field equation would be elliptic 
and thus unpredictable for d=1, too 
simple for d=2 and d=3, and unstable 
meaning unstable atoms say for d=5 or 
more. The latter point 
goes back to Ehrenfest in 1917 
\cite{Ehrenfest}, who argued that neither 
atoms nor planatary systems could be 
stable in more than four space time 
dimensions.
  
Also known is  the story that  in 
say two spatial dimensions (corresponding 
to 3=2+1 space time dimensions) an animal 
- as ourselves - having an intestine 
channel would fall apart into two pieces.
Thus by an antropic principle 3=2+1 
should not be possible to have us.

According to a review of anthropic
questions by Gordon Kane \cite{Kane}:

``One aspect of our universe we want to 
understand is the fact that we live in 
three space dimensions. There is an 
anthropic explanation. It was realized 
about a century ago\cite{Ehrenfest} that 
planetary orbits are not stable in 
four or more 
space dimensions, so planets would not 
orbit a sun long enough for life to
 originate. For the same reason atoms are 
not stable in four or more space 
dimensions. And in two or one space 
dimensions there can be neither blood 
flow nor large numbers of neuron 
connections. Thus interesting life can 
only 
exist in three dimensions. Alternatively, 
it may be that we can derive the 
fact that we live in three dimensions, 
because the unique ground state of the 
relevant string theory turns out to have 
three large dimensions (plus perhaps 
some small ones we are not normally aware 
of). Or string theory may have many 
states with three space dimensions, and 
all of them may give universes that 
contain life''.

Further  one has considered the 
renormalizability of quantum field 
theories 
not being possible for higher than 
4 dimensions, except for the scalar 
$\phi^3$ coupling theory, which is 
anyway not good \cite{Baez}.      

In theories, which like string theories
or Norma Mankoc Borstnik's model 
\cite{Norma} are Kaluza-Klein-like, the 
question 
of understanding the effective dimension 
for long distances being 3 space plus 
1 time dimension would a priori mean an 
understanding of why precisely there is 
that  number of extra dimensions being 
somehow ``compactified '' that just 
three space
dimensions survive as essentially flat 
and extended. In super-string theory the 
consistency requires fundamentally 9 +1 
dimensions of space time. If one takes it 
that it is needed that the compact space 
described by the as extra dimensions 
appearing dimensions must be a Callabi Yau
space, then since the latter has 6 
dimensions the observed or flat dimensions
must make up (9+1) - (6+0) = 3+1. 
So the combination of the susperstring 
with the requirement of using Callaby-Yau
compactification do indeed explain why
we have just the experimental number of 
dimensions\cite{Candelas1, Candelas2}.

In \cite{suext} you find:

``Now to make contact with our 4-dimensional world we need to compactify the 10-dimensional superstring theory on a 6-dimensional compact manifold. Needless to say, the Kaluza Klein picture described above becomes a bit more complicated. One way could simply be to put the extra 6 dimensions on 6 circles, which is just a 6-dimensional Torus. As it turns out this would preserve too much supersymmetry. It is believed that some supersymmetry exists in our 4-dimensional world at an energy scale above 1 TeV (this is the focus of much of the current and future research at the highest energy accelerators around the world!). To preserve the minimal amount of supersymmetry, N=1 in 4 dimensions, we need to compactify on a special kind of 6-manifold called a Calabi-Yau manifold. ''


\end{document}